\documentclass{elsart}

\usepackage{natbib}

\usepackage{epsfig}

\usepackage{amssymb}

\def\barecite#1{\citealt{#1}}
\def\refcite#1{\citet{#1}}

\def\referee#1{{#1}}

\begin{document}
%\warningoverprint{Draft v. \today}

\begin{frontmatter}

% Title, authors and addresses

% use the thanksref command within \title, \author or \address for footnotes;
% use the corauthref command within \author for corresponding author footnotes;
% use the ead command for the email address,
% and the form \ead[url] for the home page:
% \title{Title\thanksref{label1}}
% \thanks[label1]{}
% \author{Name\corauthref{cor1}\thanksref{label2}}
% \ead{email address}
% \ead[url]{home page}
% \thanks[label2]{}
% \corauth[cor1]{}
% \address{Address\thanksref{label3}}
% \thanks[label3]{}

% Potential reviewers:
% 
% Hugh Hudson (hhudson@ssl.berkeley.edu)
% Lidia van Driel (Lidia.vanDriel@obspm.fr)
% Mike Wheatland (wheat@physics.usyd.edu.au)
% 

\title{Driving Major Solar Flares and Eruptions: \\A Review}

% use optional labels to link authors explicitly to addresses:
% \author[label1,label2]{}
% \address[label1]{}
% \address[label2]{}

\author{Carolus J. Schrijver}
\address{Lockheed Martin Adv.\ Techn.\ Center, 3251 Hanover Street, Palo Alto, CA 94304}

\begin{abstract}
This review focuses on the processes that energize and trigger M- and
X-class solar flares and associated flux-rope destabilizations.
Numerical modeling of specific solar regions is hampered by uncertain
coronal-field reconstructions and by poorly understood magnetic
reconnection; these limitations result in uncertain estimates of field
topology, energy, and helicity. The primary advances in understanding
field destabilizations therefore come from the combination of generic
numerical experiments with interpretation of sets of observations.
These suggest a critical role for the emergence of twisted flux ropes
into pre-existing strong field for many, if not all, of the active
regions that produce M- or X-class flares.  The flux and internal
twist of the emerging ropes appear to play as important a role in
determining whether an eruption will develop predominantly as flare,
confined eruption, or CME, as do the properties of the embedding
field.  Based on reviewed literature, I outline a scenario for major
flares and eruptions that combines flux-rope emergence, mass draining,
near-surface reconnection, and the interaction with the surrounding
field. Whether deterministic forecasting is in principle possible
remains to be seen: to date no reliable such forecasts can be
made. Large-sample studies based on long-duration, comprehensive
observations of active regions from their emergence through their
flaring phase are needed to help us better understand these complex
phenomena.
\end{abstract}

\begin{keyword}
% keywords here, in the form: keyword \sep keyword
Sun: flares \sep Sun: magnetic field \sep Sun: emerging flux
% PACS codes here, in the form: \PACS code \sep code

\end{keyword}

\end{frontmatter}

\parindent=0.5 cm

% main text
\section{Introduction}\label{sec:introduction}
We have known of the phenomenon called 'solar flare' ever since the
first observation by Carrington and Hodgson in 1859. We have learned a
great deal in the century and a half that have passed since, but
these basic questions remain with us: What powers a solar eruption? 
What triggers it?  These questions continue to occupy us as we bring
ever-increasing instrumental and computational capabilities to bear on
the problem.  At the time of this writing, in the summer of 2008, a
search with the Astrophysics Data System yielded over 11,000 refereed
publications with the words 'flare' and 'Sun' or 'solar' in the
abstract. Even the limited focus of this paper on the energy source
and triggering process of active-region flares and their associated
eruptions is associated with thousands of studies. 

This review focuses on active-region M- and X-class flares and the
flux-rope destabilizations associated with them. I present my summary
first, followed by a synthesis scenario for the driving of field
eruptions. These two sections are meant to give the reader a context
to assess the supporting evidence that is discussed in detail,
\referee{with references}, in
subsequent sections. \referee{Readers who would rather see the more 
traditional review in which the discussions and conclusions follow the
arguments may skip Sect.~2 initially, returning to it just prior to
Sect.~13.} 

A selection of characteristic flares and
eruptions discussed in the text is compiled in Table~1. \referee{These
flares are all well-observed examples that illustrate the findings
described in the indicated parts of the text, and many are the basis
of studies that are discussed in this review.}

\section{First, the conclusions:} \subsection{Assessing key findings
and inferences}
\label{sec:questions}\label{sec:conclusion}\label{sec:discussion} One
would like to have an understanding of solar flares and eruptions that
is founded on theory and on numerical models of the behavior of
magnetic fields from before they intrude into the solar atmosphere to
their subsequent evolution within it. Unfortunately, our modeling
abilities in these respects are limited by the enormous gradients and
scale ranges involved \referee{(Sect.~5)}. For one thing, simulations of
the emergence of magnetic flux from deep within the convective
envelope \referee{(Sect.~10)} have made substantial progress in recent
years, but they all require twists of flux ropes that are not
obviously compatible with the bulk of the regions observed to emerge
through the photosphere: most solar regions emerge as bipolar regions
with fibril and loop connections through chromosphere and corona more
or less parallel to the line connecting the centers of gravity of the
two polarities, whereas simulations would have the initial fields come
up very nearly at right angles to the flux rope's subsurface direction
\referee{(Sect.~10)}.

Even when the field is in the corona, we have to be very careful in drawing 
conclusions from modeled field
topologies and derived energy estimates.
The main reasons for that problem are (1) the forces acting on the field
within the photosphere, (2) the uncertainties on vector-field
measurements, particularly on the transverse component, and (3) the large
domain that needs to be modeled to capture the connections of an active
region to its surroundings (see Sects.~\ref{sec:topology} and~8).
\referee{Observationally, pre-to-post flare changes in the photosphere are
detectable but not very strong even for the largest flares (Sect.~8.1),
while interpreting changes in the corona suffers from the fact that it
is difficult to disentangle coronal field changes from changed
thermal conditions that cause different field regions to show up
(Sect.~8.2).}

In view of these problems, most of our knowledge about what powers and
triggers flares comes from observations (supported by generic models
and theory; \referee{Sects. 7, 10, and 11}).  It appears that flares and
eruptions are intrinsic to the emergence of magnetic field interacting
with the overlying field both within and above the active region core
field \referee{(Sects.~7 and 9)}.  Hence, it is crucial that we extend our
thinking from instantaneous, 2D observations and sketches to an
intrinsically-dynamic, 3D view 
\referee{to capture the evolution of twisted field as
it emerges through the photosphere, already carrying current and
helicity well before doing so. 
Such thinking  --~clearly included in 
the sketches in
Figs.~\ref{fig:flarefluxmodel}, \ref{fig:ishiikurokawa},
and~\ref{fig:schmieder}, and further discussed in
Sect.~10~-- has} to elucidate
why the large-scale polarity-inversion lines \referee{of pre-existing
photospheric flux are such preferred sites for subsequent flux-rope
emergence}, which probably requires understanding of the deep
active-region field configuration.

Flares, flux-rope eruptions (with or without chromospheric material
showing as filaments), and CMEs associated with active-region events
are largely-overlapping populations of closely-related features of
field destabilization \referee{(Sect.~3)}.  It appears that the factors
that determine which of these features dominates the evolution are
matters of scale, available energy, twist, \referee{and} the surrounding
magnetic field \referee{(Sect.~9)}. Large statistical studies are needed
to better elucidate the relationships between these features, and to
separate the causally related parameters from the many others that
exhibit substantial correlations.

Essential to the occurrence of M- or X-class flares is a
strong-gradient polarity-inversion line (SPIL) in the strong-field
interior of an active region. \referee{I have found no cases of 
major flares} in the absence of
such a SPIL, or when there is not enough flux associated with it
(Sect.~\ref{sec:patterns}).  Such SPILs are generally associated with
flux emergence, and for the cases for which vector-magnetic data are
available, that emergence pattern looks like a wound flux-rope that
breaches the surface (see the well-studied example in the bottom panel
of Fig.~\ref{fig:flarefluxmodel}). Such systems either flare 
\referee{(often repeatedly, see Sects.~4 and~6)} or relax
back to a near-potential state
within a day or so after flux emergence ends \referee{(Sect.~9)}.

The emergence of these flux ropes requires that most of the mass that
they carry is drained back into the depths below the photosphere. For
a spiraling, undulating field this presents problems for the
field-line segments that thread the corona on two sides with a
sub-photospheric segment in \referee{between: these undulating field lines
(Sect.~\ref{sec:reconnection}) form anchors to the rising flux ropes
because the slumping mass remains tied to the magnetic field.} The
pinching-off of the mass-carrying pockets by near-photospheric
reconnection may be one of the mechanisms involved in the
destabilization of the field configuration.

Even as flux-rope emergence appears fundamental to the driving of
major field destabilizations, it does not necessarily do so.  Perhaps
the available energy can be dissipated in a series of smaller flares
(e.g., \barecite{schrijver+etal2005a}) --~\referee{although the flare
magnitude spectrum per region appears statistically to be the same
(but see Sect.~4)}~-- or by enhanced coronal heating (e.g.,
\barecite{falconer+etal1997}) --~\referee{although the coronal energy
losses appear to be independent of the properties of the
vector-magnetic field in the photosphere (e.g.,
\barecite{fisher+etal98})}. Or, perhaps this happens because the
twist of emerging flux ropes must be aligned with the direction of the
overlying field to build up energy; if largely anti-parallel,
reconnection may happen immediately without the buildup of much energy
(cf., simulations by \barecite{fan+gibson2004}). More comprehensive
multi-day studies of active regions are critical to understand this.

The finding that the flare-magnitude power-law spectrum of individual
active regions is similar to the ensemble's average distribution (see
Sect.~4) suggests that the energy-release process is the result of
some process of self-organization of the coronal field. But it may
also be a consequence of the sub-photospheric flux-rope shredding
\referee{(Sect.~10)}.
Repeated flaring of a region with continued, intermittent flux
emergence supports the latter, while observations that a flare or
eruption releases only part of the region's non-potentiality supports
the former. Understanding the consequences of these processes on the
\referee{possible existence} of a deterministic flare forecast requires systematic
large-sample studies to differentiate the roles of the
sub-photospheric and atmospheric processes.

With respect to flare forecasting \referee{(see
Sect.~\ref{sec:instability})}, I note that (1) if there is no sign of
substantial flux ropes within the photosphere for about a day, no
major flares will occur within the next few hours, but (2) flux
rope-emergence may lead to major flares within about a quarter of a
day, and (3) may be associated with one or more major flares, or with
a series of smaller flares, or with gradual dissipation of the
flux-rope's energy. At present, there are no flare-forecast metrics
with an outstanding 'skill score', although the total flux near a SPIL
appears to be a good indicator of the flare potential of an active
region. It remains to be seen whether the limited skill score of
forecasts can be improved with additional observations, or whether the
flare process is intrinsically probabilistic with the possibility of a
wide range of responses to any given boundary perturbation.

\subsection{A scenario for initiation of major flares and eruptions}\label{sec:scenario}
The literature on flares and field destabilizations, on properties of
flux emergence, and on field topology and reconnection, suggests the
following scenario for the majority of the large active-region flares
that are the focus of this review, \referee{and whose component elements
are discussed in subsequent sections. The top panel of Fig.~1 
illustrates the components of this scenario.} 

\begin{figure}
\begin{center}
\includegraphics*[width=\textwidth]{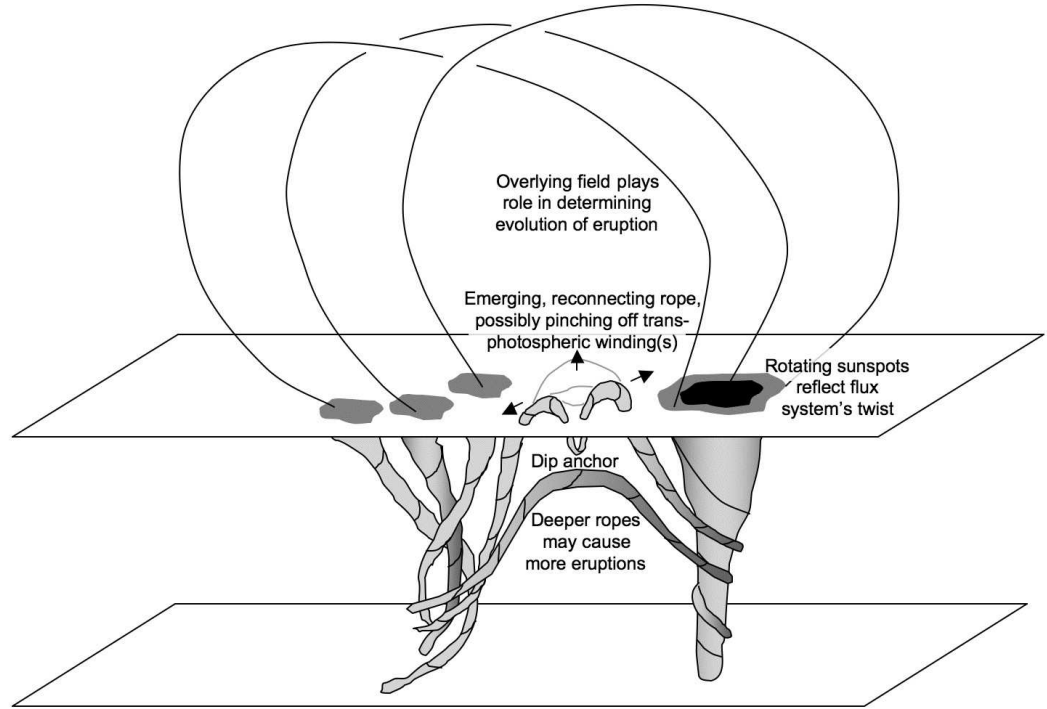}

\includegraphics*[width=\textwidth]{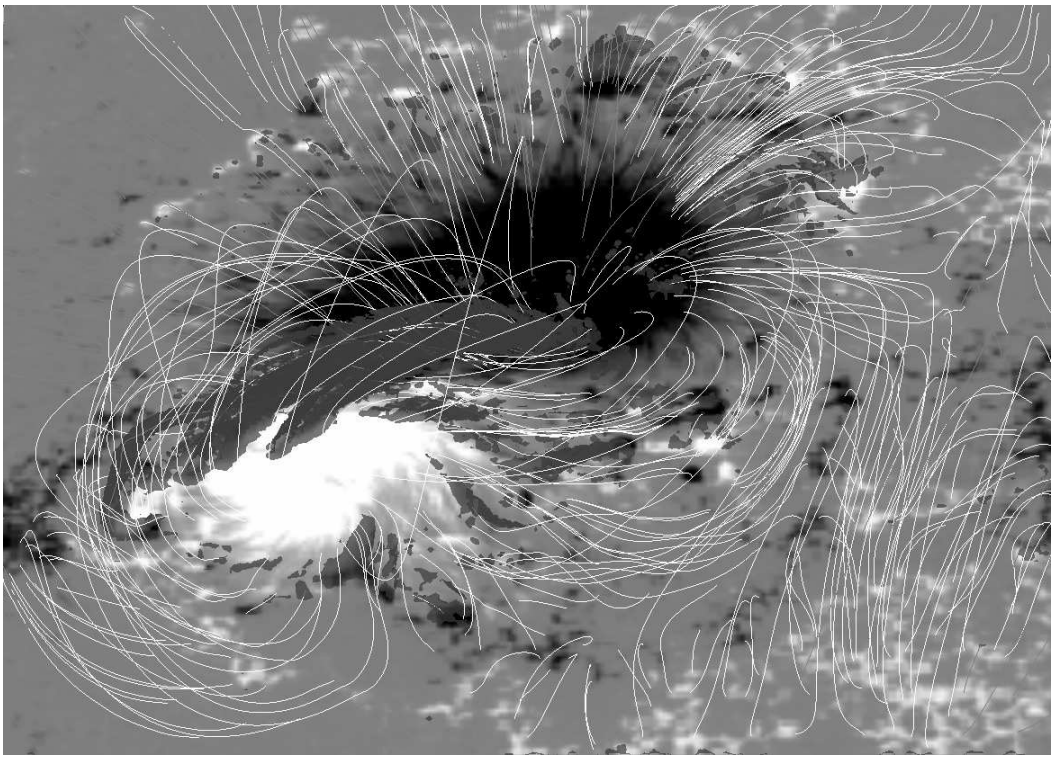}
\end{center}
\caption{\label{fig:flarefluxmodel}
{\em Top:\/} Schematic representation for an emerging field
configuration involved in flaring or flux-rope eruption as discussed in
Sect.~\ref{sec:scenario}.  {\em Bottom:\/} NLFF field model over an
active region prior to an X3.4 flare (case 8 in Table~1), rendered
from a perspective 50 degrees from the normal view.  The two main
spots are connected by a flux rope with strong 
electrical currents (shown by
a grey isosurface) within which field lines spiral
from end to end on the photosphere. The Hinode vector-magnetic data and the
extensive NLFFF modeling are discussed by Schrijver et al.\
(2008). \referee{See Sect.~5 for a discussion of NLFFF modeling capabilities.}}
\end{figure}\nocite{schrijver+etal2007}
First, a large active region \referee{emerges}, 
itself already carrying substantial net
twist, helicity, and electrical currents, as witnessed by surface
motions such as rotating sunspots, by its net kinetic helicity, and by the
generally complex shape of the magnetic field.

Subsequently, a current-carrying bundle of magnetic flux (either a
single flux rope, or an ensemble of such ropes, perhaps formed by
pre-emergence fragmentation) breaches the solar surface
(see Fig.~\ref{fig:flarefluxmodel}-top). This rope may
be a separate rise of flux through the convective envelope, or may be
a late emergence of strands of flux from the same parent flux bundle
that formed the active region. One flux rope, or a series of them, may
emerge, forming a strong-gradient polarity-inversion line (SPIL)
characterized by two parallel opposite-polarity ridges of
line-of-sight flux in close proximity.  At high resolution, the vector
magnetic field across the SPIL will be highly sheared, and rotating as
the flux bundle moves through the photosphere.

Interaction with the near-surface convection, radiative cooling in the
stratified atmosphere, and the draining of heavy plasma into
surface-crossing undulations or windings of the field appear to result
in a number of 'anchor' points for the emerging flux
rope(s). Near-surface reconnection (``tether cutting'') enables the
rope to break any ties to the interior between its two extremities, so
that part or all of it can rise relatively freely into the overlying
corona.

\referee{Models lead to the  hypothesis that 
a rope may be unstable, and accelerate into an
eruption, often associated with flaring, 
if the strength of the pre-existing field decreases rapidly enough
with height (Sect.~9)}. Part of
all of the emerged strands may participate in this, depending on the
remaining sub-photospheric anchors. If the field gradient slows with height or
changes sign, the flux rope may stabilize again, resulting in a
confined flare or failed flux-rope (and, often, filament) eruption. If
the field gradient is large (particularly if the field goes through a
null, as in the ``breakout'' concept), the eruption may proceed
higher, if not in fact into the heliosphere as a CME. If the winding
of the emerging flux rope is too strong, then (confined or ejective)
kinking may happen after any photospheric anchors between the end
points are released; such kinking may, but needs not, transition to an
ejective eruption.

Subsequent emergence of a series of flux rope fragments is likely 
responsible for repeating and homologous flares, which can manifest
themselves as leaving non-potential energy (or helicity, or
electrical currents, or filaments, \ldots) behind in the corona if these
are not (yet) unstable or released from their photospheric anchors. Or
they may manifest themselves as repeating flares when subsequent
strands of the flux rope emerge. 

How general is this scenario? The following sections address various
aspects of this, but let me point out here, in summary, that the vast
majority of major flares are associated with flux emergence (within
the preceding day or so) in the form of tightly wound flux ropes in
active regions with large-scale twist, that field gradients above the
zone near the polarity-inversion line help differentiate confined from
ejective filament eruptions, while the photospheric differences
between immediately before and after the flare are generally
\referee{so subtle they are detected only for the largest flares that
have been observed at high cadence and then only after careful
inspection (Sect.~8.1).} Conversely, no major flares occur in regions
that have a largely potential appearance, or more than about a day
after the emergence of substantial flux ropes \referee{(Sect.~9)}.  It
remains unclear what role tether-cutting or anchor-cutting play in
near-photospheric reconnection \referee{(Sect.~11)}. These processes
have both observational and theoretical support, but it remains
unclear how instrumental they are in the triggering of flares, either
by themselves or in combination with the (evolving) field gradient
above the polarity-inversion line \referee{(Sect.~9)}.

\section{Flares and eruptions: confined, failed, ejective}\label{sec:types}
Flares, filament eruptions, and coronal mass ejections (CMEs) are
three aspects that manifest themselves following the destabilization
of the coronal magnetic field. Their precise relationship remains an
area of intense study, which is aided by the ever-increasing
observational coverage of the Sun and inner heliosphere, yet hampered
by the fact that different observatories are involved that generally
do not have overlapping fields of view, and in many cases do not have
continuous coverage of a given target active region.  Another problem
is, of course, that the sources of CMEs that originate on the far side
of the Sun cannot be observed, \referee{while there is
ambiguity in 
differentiating front-side CMEs from far-side CMEs} --~at least there was
until the launch of the STEREO spacecraft late in 2006.

In addition to flares and CMEs, there are the 'failed eruptions' in
which filaments begin to rise, but then stop their progress while
twisting and writhing \referee{(see, e.g., \barecite{gilbert+etal2007},
for a discussion of this phenomenon and further references)} within
the confines of what appears to be a domain of connectivity for the
field bounded by a separatrix; some particularly well-observed
examples of this with TRACE  are cases $1-7$ in
Table~\ref{tab:regions}.  These failed eruptions are associated with
flares up to low M class. At least some of these failed eruptions have
no CME counterpart (e.g., \barecite{ji+etal2003}).

\begin{table}[t]
\begin{centering}
\caption{\label{tab:regions}
Sample flaring regions and flare events discussed in the text.}
{\scriptsize
\begin{tabular}{lllr}
\hline
No. & Time, flare magnitude & Notes & Sect. \\
\hline
1& 2003/05/02 02:47 M1.0 & Apparently failed filament eruption obs.\ by TRACE/171\AA. & \ref{sec:types}\\
2& 2002/05/27 18:00 M2.0 & Apparently failed filament eruption obs.\ by TRACE/195\AA.&\ref{sec:types}\\
3& 2001/08/01 18:30 ---&Apparently failed filament eruption obs.\ by TRACE/171\AA.&\ref{sec:types}\\
4& 2001/04/15 21:59 C5.1 &Apparently failed filament eruption obs.\ by TRACE/171\AA.&\ref{sec:types}\\
5& 2000/11/05 00:17 C5.4 &Apparently failed filament eruption obs.\ by TRACE/171\AA.&\ref{sec:types}\\
6& 1999/10/20 05:53 M1.7 &Apparently failed filament eruption obs.\ by TRACE/171\AA.&\ref{sec:types}\\
7& 1999/05/31 22:10 ---&Apparently failed filament eruption obs.\ by TRACE/171\AA.&\ref{sec:types}\\

8& 2006/12/13 02:40 X3.4& Extensive NLFFF modeling by Schrijver et al.\ (2008).&\ref{sec:topology}\\

9 & 2008/08/25 16:23, X5.3 & Ribbons distant from main PIL within AR complex. & \ref{sec:topology}\\
10 & 2004/07/16 13:49, X3.6 & Ribbons distant from main PIL within AR complex. & \ref{sec:topology}\\
11 & 2001/04/10 05:06, X2.3 & Ribbons distant from main PIL within AR complex. & \ref{sec:topology}\\
12 & 2001/10/19 00:47, X1.6 & Ribbons distant from main PIL within AR complex. & \ref{sec:topology}\\
13 & 2003/03/18 11:51, X1.5 & Ribbons distant from main PIL within AR complex. & \ref{sec:topology}\\
14 & 2000/06/07 15:34, X1.2 & Ribbons distant from main PIL within AR complex. & \ref{sec:topology}\\
15 & 2004/07/16 10:32, X1.1 & Ribbons distant from main PIL within AR complex. & \ref{sec:topology}\\
16 & 2002/08/21 05:28, X1.0 & Ribbons distant from main PIL within AR complex. & \ref{sec:topology}\\
17 & 2004/07/17 07:51, X1.0 & Ribbons distant from main PIL within AR complex. & \ref{sec:topology}\\

18 & 2001/06/15T10:01 M6.3 & Ribbons in adjacent AR at $\sim $5\,arcmin. & \ref{sec:topology}\\
19 & 2002/03/14T01:38 M5.7 & Ribbons in adjacent AR at $\sim $2\,arcmin. & \ref{sec:topology}\\

20 & 2002/07/31T01:39 M1.2 & Sympathetic flare in AR 4\,arcmin.\ away. & \ref{sec:topology}\\

21 & 2005/09/07-17 10X, 24M & AR 10808: 10 X-class and 24 M-class flares. & \ref{sec:homology}\\
22 & 2003/10/18-11/05 11X & AR 10486: 11 X-class flares, including a record X28. & \ref{sec:homology}\\
23 & 2000/11/24-27 5X, 2M & AR 9236: 5X and at least 2 M-class flares. & \ref{sec:homology}\\

24 & 1999/09/19 08:40 & Post-event arcade above a residual filament (TRACE/171 \AA)&\ref{sec:homology}\\

25 & 2000/07/19 23:39 ----&Erupting filament with decreasing twist (TRACE/171\AA)&\ref{sec:beforeafterabove}\\

\hline
\end{tabular}
See http://trace.lmsal.com/POD/bigmovies/ for a collection of TRACE movies. 
}
\end{centering}
\end{table}

The fraction of flares that is associated with CMEs increases rapidly
as we go from small flares to large X-class flares, reaching close
to 100\%\ for the largest ones.
For example, \refcite{wang+zhang2007} find that
90\%\ of 104 studied X-class flares are associated with eruptions, and
\refcite{yashiro+etal2005} find that the fraction of flares
associated with CMEs increases from 20\%\ for C3-9 flares to 100\% for
flares above X3. \refcite{andrews2003} finds that 40\%\ of M-class flares
and all X-class flares 
in his sample are associated with CMEs.
 
It remains unclear precisely what determines whether a large flare is
associated with a successful ejection or with a failed eruption. One
clue is found in a study by \refcite{wang+zhang2007}, who find that
the 10\%\ of X-class flares that they find not to be associated with
CMEs tend to occur closer to the magnetic
center of gravity of the region, leading them to suggest that 
eruptive flares lie closer to the periphery of
the active regions. Another distinction, \referee{suggested} by
\refcite{nindos+andrews2004}, \referee{may be} the field's helicity or
net electrical current: they find that the best-fit linear force-free
(LFF) fields have a value of the twist parameter $\alpha$ that is
statistically larger in CME-producing flaring regions than in those in
which the flares are not associated with CMEs, although there is
substantial overlap in the samples. \referee{It is not clear what such
LFF fits measure, though; a NLFFF formalism would have been appropriate,
but that, too, is fraught with problems (Sect.~5).}

Larger (brighter and often hotter) flares tend to be associated with
faster and/or wider CMEs (which presumably tend to be more energetic);
see, for example, \refcite{kay+etal2003},
\refcite{yashiro+etal2005}, \refcite{guo+etal2007}, and
\refcite{georgoulis2008}. Moreover, the fastest CMEs originate in 
active regions. For example,
\refcite{wang+zhang2008} find that 90\%\ of the fastest 0.5\%\ of CMEs 
(with speeds above $1500$\,km/s) originate in active regions.
This suggests that the energy in the surrounding magnetic field is involved in
determining whether a filament eruption develops into an
ejection. Other aspects are the geometry and gradient of
the overlying field --~see Sects.~\ref{sec:statistics} 
and~\ref{sec:overlying}.

\section{Flare statistics from large to small}\label{sec:statistics}
The energy that is released during flares is converted into non-thermal
particle populations, thermal populations, and bulk kinetic motion in
case of ejecta. Although the initial release partitions
the energy  over the 
thermal and non-thermal particle populations (see, e.g., 
\barecite{krucker2005}, \barecite{sainthilaire+benz2005}, 
and references therein), much of the non-thermal population is
eventually converted into thermal energy and into the radiation that
we see as the X-ray flare.  Measurements of the radiative losses
during flares (using, e.g., peak fluxes or fluences) thus capture much
of the flare's energy release, apart from the bulk kinetic energy in
ejections and the escaping particle populations.

\begin{figure}
\begin{center}
\includegraphics*[width=0.6\textwidth,viewport=40 90 500 650,clip]{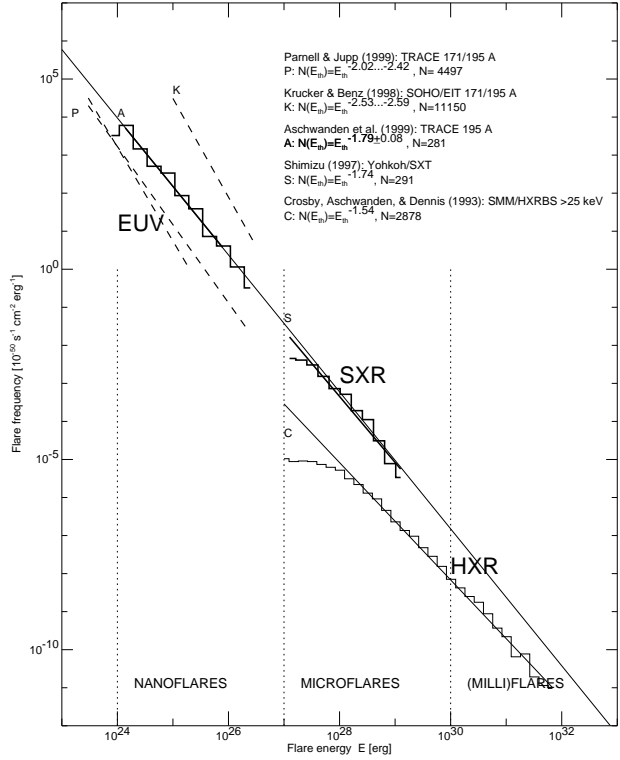}
\end{center}
\caption{\label{fig:powerlaws}
Distribution functions for estimated flare energies derived from observations
in the extreme ultraviolet, soft X-rays, and hard X-rays. The energy
range in this diagram spans eight orders of magnitude.  From
Aschwanden et al.\ (2000).}
\end{figure}

Despite this incomplete access to the total energy
involved in flares and eruptions, many studies suggest that
peak fluxes or fluences are among the useful metrics for flaring activity.
Distribution functions of estimated energies of solar flares
can be approximated by power laws rather well (see, e.g., the
overview in Table\,I in
\barecite{aschwanden+etal1998}, who review flare statistics from
hard X-rays to radio as well as energetic particle events; see also 
\barecite{su+etal2006}, for an analysis of RHESSI flares; and
\barecite{hudson2007}, for a compilation of 32 years of GOES $1-8$\AA\ flares).
Figure~\ref{fig:powerlaws} summarizes the flare-energy power laws
from nano- to X-class flares (with energies of ${\cal O}(10^{31-32})$\,ergs). 

The existence of power laws in not limited to
flares. \refcite{robbrecht2007} finds essentially a power-law
distribution for the widths of CMEs.  Flares with or without CMEs may
be statistically different:
\refcite{yashiro+etal2006}, for example, \referee{suggest} that power-law
distributions for peak fluxes, fluences, and duration are
significantly steeper for flares without CMEs than for flares
associated with CMEs. On the other hand, 
\refcite{wheatland2003} reports that the waiting-time distribution
for CMEs is like that of a time-dependent Poisson process, 
similar to that found for flares observed in the GOES records.

\nocite{aschwanden+etal2000b}
In reviewing the evidence for the common occurrence of power laws for
flare properties, \refcite{charbonneau+etal2001} conclude that their
existence ``suggests that the flaring process is intrinsic to coronal
magnetic fields, even though the flaring rate may be controlled by
extrinsic factors, such as magnetic flux emergence in the
photosphere.'' 
The inference that the power-law distribution of flare properties
may reflect some intrinsic property of the corona is consistent
with the finding that the same power-law distributions are found
when looking at subsets of flares from particular regions:
\refcite{wheatland2000b} shows that the distributions of peak flare
strengths for individual active regions is consistent with the joint
pdf for all regions, and concludes that, apart from the frequency,
'the statistics of flares appear to be independent of the physical
characteristics of the originating active regions', at least within
the statistical uncertainties. \referee{A study by \refcite{kucera+etal1997}
supported an upper-limit to flare magnitudes dependent on the region's
spot coverage (cf., Sect.~7). They  also found a reduced number
of small flares in regions with high spot coverage,
but suggest that an observational
bias in their data requires further study on that issue; 
\refcite{wheatland2000b} finds no significant evidence for such an effect.}

\refcite{hudson2007} comments on the predictability of
X-class flares from a cycle perspective. He points out that the
frequency of these events through three solar cycles does not follow
the Poisson statistics based on metrics for the sunspot cycle that
appear applicable to lesser flares: there are more major flares
in the cycle-minimum phases than there ought to be if their occurrence
scaled simply with active-region frequency. The flaring frequency
appears rather
specific to active region properties and their
sub-photospheric sources, and these appear to change with the cycle in
ways that remain to be uncovered.

The statistical commonality of the shape (apart from the overall
frequency) of the flare-energy spectrum for all regions has led to the
introduction of the concept of the 'avalanche model' or
'self-organized criticality' (SOC; see, e.g., the review by
\barecite{charbonneau+etal2001}) in which the size of a flare is only
statistically determined, and in which the flare's energy is always
much less than the available total energy (thus allowing repeating and
homologous flares with little change in the boundary field, see
Sect.~\ref{sec:topologyribbons}). The SOC of the coronal field may
develop as a consequence of the processes that inject the
non-potential energy into the active-region coronae -~discussed
further in the following sections. We have to wonder, though, whether
the 'self-organization' reflects a coronal property or is rather a
property of the subsurface processes that supply the energy in the
first place (see Sect.~10). With that in mind, let me now proceed with
a discussion of what we know -~or rather what we do not know~- about
flare energetics based on field measurements.

\section{Field models and topology, free energy, and helicity}
\label{sec:topology}
Models of solar flares would have free energy in excess of a
lower-energy field to which it can relax,
(relative) helicity, and field topology play important
roles in determining whether a flare can occur, where it will begin
its development, and what the final state of the coronal magnetic field
can be. In order to derive such properties, ideally one would like
to model the coronal field in detail based on the observable
boundary conditions in the low atmosphere.

The electrical currents in the solar atmospheres above flare-prone
active regions are substantial, and we know that such currents can
lead to significant changes in field geometry (see
Sect.~\ref{sec:overlying}) and topology (e.g.,
\barecite{hudson+wheatland99}).
Unfortunately, modeling the coronal field has turned out to be
exceedingly difficult. Problems abound (see, e.g.,
\barecite{klimchuk+etal1992},
\barecite{mcclymont+etal1997}, 
\barecite{schrijver+etal2005b},
\barecite{metcalf+etal2006}, \barecite{metcalf+etal2007};
\barecite{schrijver+etal2007}, and \barecite{derosa+etal2008}): 
nonlinear force-free field (NLFFF) models or virial-theorem
applications based on photospheric vector fields (that are subject to
problems of their own, including inversion problems and an intrinsic
180$^\circ$ ambiguity for the transverse field) have to be developed
further if they are to provide reliable estimates of the free energy,
relative helicity, or topology. \refcite{schrijver+etal2007}
illustrate this very clearly.  They compare field models for AR~10930
(case no.~8 in Table~\ref{tab:regions}) based on high-quality
Hinode/SP vector-magnetic measurements, using a variety of NLFFF
algorithms and a range of boundary conditions. Only one of these
solutions approximates the observed coronal configuration
\referee{(bottom panel in Fig.~1)}. A
subsequent study by \refcite{derosa+etal2008} based on
Hinode and STEREO observations concludes that there is, as yet, no
successful strategy for NLFFF modeling that can be applied under
general circumstances to yield significant energy estimates and
reliable field geometries. Consequently, in my opinion, we have yet to
succeed to a level that makes comparison of observation-based field
models with theoretical concepts valuable to differentiate between
different flare scenarios.

\referee{When} studying the changes in
field properties from before to after a flare, an added problem to the
one discussed above is that the volumes that can be effectively
modeled within computational limitations are so small that 
\referee{they have no real solar counterparts that may be considered to be
closed to any physical quantity}. Hence, the apparent convenience
of having a conserved quantity -~the relative helicity~- can in
reality not be exploited effectively, even if it could be reliably
measured.

Then there is, of course, the problem that
the coronal field is anchored in the photosphere (which shows
only subtle lasting changes during the course of a flare, see
Sect.~\ref{sec:beforeafter}), which limits the field's relaxation
towards a LFF field. Moreover, in any given flare, only part of
a region's helicity and energy may be involved (see Sect.~7). 
Consequently, even if
we were able to measure the 'free energy' in excess of either a
potential field or a LFF field (at the same or any other 
relative helicity), we would
only be able to place an upper limit on the magnitude of a flare.

Those studies that do try to measure energy or helicity in spite of
the above problems, typically find that the energy or relative
helicity in the region suffices to power a flare or CME, and that they
typically decrease from before to after the flare, even
when measuring relative to a LFF field at the same helicity
(e.g., \barecite{bleybel+etal2002}; 
\barecite{metcalf+etal2005};
\barecite{regnier+canfield2005};
\barecite{regnier+priest2007} [also \barecite{regnier+priest2007a}]; 
\barecite{bobra+etal2007};
\barecite{liu2007}). Perhaps we can have more confidence in relative
measurements than in absolute measurements? Or perhaps we are
establishing that flares release only a small fraction of the total
energy associated with the field, so that we should not be surprised
that the energy in a (N)LFF field model exceeds the potential or
LFF-field energy by enough to let us conclude that there is enough
energy to power the flare. That would be a rather more spurious
result than one would like to have.

The application of the virial theorem to the lower-boundary field
suffers from its own problems. For example,
\refcite{wheatland+metcalf2006} argue that the virial estimates of the free
energy in active region fields based on the surface vector field can
be somewhat improved over the simple boundary where that ignores
Lorentz forces, but that the results are subject to very substantial
uncertainties. An application of the virial theorem to a model field
in the detailed study by \refcite{metcalf+etal2007}, for example, shows that
the magnitude of the uncertainty in the virial-theorem estimate of
free energy is comparable to that in NLFF models. 

If we cannot, yet, rely on measurements based on field models, can we
at least rely on the overall geometry of the field, and in particular
properties of its skeleton, such as (quasi-)separatrices?  For nearly
potential fields, this should be possible, but the largest flares and
CMEs are associated with decidedly non-potential fields (see
Sect.~\ref{sec:overlying}), which unfortunately cannot be successfully
modeled. For example, \refcite{hudson+wheatland99} point out that the
observation that flares occur primarily at separatrices (or
quasi-separatrix layers) is largely unproven because of the mapping
ambiguities in 3D field models. Where such agreements are found, we
should probably even question whether that is because in such cases
the non-potentiality (or force-free parameter $\alpha$) is rather
small, so that we are largely looking at the properties of a
potential-field configuration.

\label{sec:topologyribbons}
One property of
flares that has proven quite telling as far as the reconnected field
involved is the phenomenon of flare ribbons. They have long been
associated with impact sites of energetic particles caused by
reconnection (for one detailed study on this, see
\barecite{asai+etal2003}).
Generally, flare ribbons consist of partially- or fully-outlined
fronts moving away on either side of a pronounced polarity inversion
line; see, e.g.,
\barecite{fletcher+hudson2001}, and \barecite{fletcher+etal2004}, for 
discussions of particular events, with references to other studies).
They are interpreted as the signatures of reconnection following
the eruption of field
through an arcade-like overlying field, 
leading to particle acceleration and precipitation,
energy thermalization, and thus eventually to the
two-ribbon flare observable in visible light, UV, and
EUV. In the case of a true ejection of a (frequently filament-carrying)
flux rope into interplanetary space, one would expect that the
erupting field has to break through more than the lowest domain of
connectivity for the field, and that consequently more 'ribbons'
should show up than only the innermost ones (which is likely to be the 
brightest and most readily observable one). It seems that this
phenomenon of more distant particle impact sites has received
relatively little attention, perhaps because of the problem of limited
fields of view, perhaps because their interpretation requires a
large-scale field model that is generally not available. Nevertheless,
a study of such events (of which I list some examples in
Table~\ref{tab:regions}, cases $9-19$) may prove quite telling on 
large-scale field topology and its changes, and
should be much easier with the full-Sun field of view of the future
Solar Dynamics Observatory. Studies of such larger areas may also shed
light on the phenomenon of the sympathetic flare (an example of which
is case no.~20  in Table~\ref{tab:regions}).

\refcite{li+etal1997} find that hard X-ray flare footpoints occur
preferentially at the edges of channels of strong vertical currents as
inferred from vector-magnetograms. This is suggestive of reconnection
between a current-carrying rope and its surrounding field at flare
onset (cf., Sect.~8.1).  Higher-resolution Hinode data are well suited
for a follow-up of that work, once the cycle picks up strength.

Initial flare
brightenings, whether in ribbons or in more compact impact sites or
compact flaring sites, generally lie close to the primary, strongest
polarity-inversion line: \refcite{schrijver2007} find that the
average distance $D$ of the brightest TRACE (E)UV location at flare
onset (the first brightening within about 1\,ksec from the GOES flare
start time,  often shown by the EUV diffraction pattern associated
with the filter support grid) 
to the nearest point on a SPIL within the region is
20\,Mm.  The distribution of $D$ is strongly peaked at small
distances, such that about 70\%\ of the flares have $D < 15$\,Mm,
compared to a typical scale for the 
active regions of at least 150\,Mm.  Such distance measurements can be used
in combination with detailed models, as in the
study by \refcite{gary+moore2004}, who discuss a filament eruption in
which the initial brightenings lie relatively far enough from the
polarity-inversion lines that they argue this is evidence for the
initial reconnections to lie above, rather than below the erupting
filament. I return to the role of SPILs in Sect.~\ref{sec:patterns}
and to the merits of 'breakout' and 'tether-cutting' concepts involved
in the argument of Gary and Moore in Sect.~\ref{sec:overlying}.

\section{Homology and repeaters}\label{sec:homology} 
Among the potentially telling properties of flares we find the
phenomenon of repeating flares for a given active region, and in
particular the sub-category of the homologous flare. In homologous
flares very nearly the same site and flare ribbons are involved in
sequences of two or more flares.  Among the well-studied regions of
repeated flaring, we find, for example, AR~9236, AR~10808 (\referee{among} the most
flare-productive regions recorded \referee{based on the GOES soft X-ray
data}), and AR~10486 (see entries
$21-23$ in Table~\ref{tab:regions}).

The repeating (possibly homologous) flare suggests that either not all
of the available energy is released in a flare, or that energy
continues to be supplied to the region so that a series of flares of
various magnitudes can occur. Evidence
is found for both of these possibilities.  For example,
\refcite{bleybel+etal2002} study a case in which the post-flare
configuration remains inconsistent with a LFFF model (either based on
the pre-event or the post-event relative helicity \referee{as determined
by a Grad-Rubin based NLFFF code}).
Multiple studies have reported on continued flux emergence
between successive homologous flares (see, for example,
\barecite{ranss+etal2000};
\barecite{nitta+hudson2001};
\barecite{sterling+moore2001};
\barecite{takasaki+etal2004}; and
\barecite{dun+etal2007}). Some of these repeater events clearly show
that not all the available energy is taken away in an event. One
such example is that of case 24 in Table~\ref{tab:regions}
which shows a post-eruption arcade over a residual filament structure.

Some studies argue that systematic motions, such as the persistent
rotation of sunspots, are involved in rebuilding the energy for
homologous flares (see also Sect.~\ref{sec:subsurface}), such as
\refcite{zhang+etal2008} (see also the related numerical model by
\barecite{devore+antiochos2008}). The phenomenon of the rotating
sunspot is generally associated with ongoing flux emergence, however
(such as in the case of the X17.2 flare Oct.\ 28, 2003, one of the
subjects of \barecite{zhang+etal2008}, part of entry 22 in Table~1),
and without flux emergence, major flares rarely, if at all, occur (see
Sect.~8).  Besides, it is hard to estimate whether flux emergence or
spot rotation adds most energy (e.g., \barecite{regnier+canfield2005})
in view of the discussion in Sect.~\ref{sec:topology}.  The
reformation of loop systems involved in long-distance, even
trans-equatorial connections (\barecite{khan+hudson2000}) is harder to
interpret, but perhaps also hinting that not all energy is removed
during eruptive events.

\begin{figure} \begin{center}
\includegraphics*[width=0.75\textwidth]{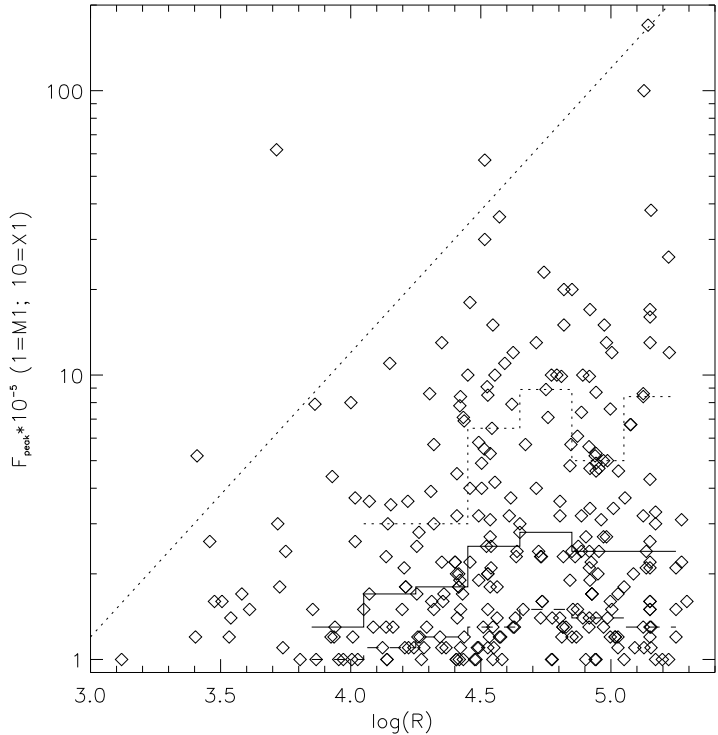} \end{center}
\caption{\label{fig:schrijver} Scatter diagram of peak flare flux
densities (GOES $1-8$\,\AA\ passband in W/m$^2$) versus the absolute
total flux near the strong-field, high-gradient polarity inversion,
$R$ (in relative units).  The histograms in this figure show the
25-th, 50-th, and 75-th percentiles of the flare distribution as
function of $R$. The dotted straight line at $F_{\rm peak}=1.2\times
10^{-8}\,R$ has 99\%\ of the observed flares below it.  The only flare
that clearly lies above this line is an X6.2 flare that started around
2001/12/13 14:30\,UT. The value of $\log{(R)}$ in this active region
decreased from 4.2 half a day prior to the flare to 3.7 at the time of
the flare, but at 4.2 would still lie well above the dotted line.
This region has a narrow fibril of one polarity embedded within an
area of strong field of the opposite polarity, which likely causes the
measure $R$ to strongly underestimate the true flux involved (see
footnote). This may (in part) explain its unusual position in this
diagram. From Schrijver (2007).}  \end{figure}\nocite{schrijver2007}
\section{Patterns in the surface field}\label{sec:patterns} The most
frequently used observable of the magnetic field involved in impulsive
and eruptive events is the surface line-of-sight field, largely
because it is readily measured (and has been, routinely, for a long
time). Increasingly, vector-magnetic measurements have become
available. Often, the frequency of either of these measurements is
rather low, in part because of observational limitations for
ground-based observatories. Consequently, many studies have focused on
finding properties of patterns of the surface magnetic field in
snapshot \referee{line-of-sight} magnetograms that might shed light on the
powering and triggering of impulsive phenomena.

One type of study focuses on statistical properties of the magnetic
field of active regions. \referee{First, it appears that M- and
X-class flares occur in sunspot-containing regions with very few
exceptions. The work of, e.g., \refcite{dodson+hedeman1970} and
\refcite{ruzdjak+etal1989} points out that whereas flares do occur in
spotless regions, these account for only about 2\%\ of all flares
while the largest documented flare in soft X-rays of this type was
classified as M1.2. This is the only documentated case I found for
which a major flare in soft X-rays occurred in a spotless region; all
of the 289 M- and X-class flares, for example, that were studied by
\refcite{schrijver2007} --~which were the basis of the study
resulting in Fig.~3~-- have spots associated
with them.}

\referee{Another statistical property was pointed by by
\refcite{canfield+russell2007}, who} find that a tessellation of
active-region magnetograms at a few arc-second resolution yields
log-normal distributions, suggesting that the flux bundles were
subjected to repeated random fragmentation during their rise to the
surface (compare the study by \barecite{bogdan85}, on a similar size
distribution of spot areas).  They find no significant correlation
between the degree of fragmentation and either the flare rate of the
active regions or the mean density of kinetic helicity, concluding
that it is more likely that large-scale helicity in the flow is
related to flaring than small-scale interactions between convection
and rising flux. On the other hand, \refcite{abramenko2005} finds that
the indices of \referee{power spectra of the photospheric
magnetic field} of flare-prone regions increase
from $-1.7$ for flare-quiet up to as steep as $-2.3$ for regions
producing high X-class flares (see her Fig.~5). It is not clear how
the preceding two results are to be reconciled.

Some of the salient features in the active-region magnetic field have
received far more attention than the above-mentioned statistical
properties, in particular the properties of the polarity-inversion
lines. Despite all this attention, no measure (either single or
combined) has been identified as being particularly well correlated
with the flare activity of active regions.  For example,
\refcite{georgoulis2008} studies 23 eruptive active regions and finds
that for active regions with 'intense polarity inversion lines' (as
measured by a metric that combines fluxes and field-line lengths from
connectivity matrices between positive and negative fluxes in
magnetograms) tend to have both stronger flares and faster
CMEs. \refcite{wang+zhang2008} find that the length and number of the
polarity-inversion lines (PILs) are good indicators of fast vs.\
slower CMEs. The findings by \refcite{falconer+etal2006} (expanding on
Falconer et al., 2002,
2003)\nocite{falconer+etal2002}\nocite{falconer+etal2003} suggest a
more complex dependency: for a sample of active regions with one
well-defined, dominant polarity-inversion line they find that several
measures are correlated similarly to the region's flare
potential. These measures include the length of the strong-shear
segment of the PIL, the length of the main PIL with a strong field
gradient across it, the total net current, and the best-fit $\alpha$
from a LFF field based on a comparison of the photospheric vector
field with the model field footprint. They suggest that the best
metric for the total magnetic free energy is, in fact, a combination
of total flux and total twist.

The ambiguity in findings in general is likely in part the result of
the correlation between multiple parameters, and disentangling these
requires large statistical samples. The study by
\refcite{leka+barnes2007} is an excellent example of that, but still
leaves us with an unclear picture. They analyze 496 active regions to
establish which parameters of the photospheric field are most
successful in differentiating flaring from non-flaring regions. They
measure 28 different properties from the vector-magnetic field
measurements and differentiate flaring from non-flaring regions within
24\,h of these magnetograph observations. Their discriminant analysis
has a success rate of 93\%\ for M1 flares or larger (compared to 91\%\
when simply labeling all regions as flare-quiet, see
Sect.~\ref{sec:instability}), or 80\%\ for C1 or larger (compared to
70\%\ in the absence of any information on the region).  They find
that the best-performing single parameter for large flares is a proxy
for the magnetic free energy in the region; other top performers are
variables that measure the field's various properties integrated over
the entire active region.

\begin{table}[t]
\begin{center}
\caption{Likelihood of X or M flares within 24\,h of the 
determination
of the unsigned magnetic flux ${\cal R}$ within about 15\,Mm of
high-gradient, strong-field polarity-separation lines (see Schrijver, 2007,
for a detailed description and normalization factor). 
\referee{The likelihoods are sampled at 0.5\,dex intervals, but
vary smoothly with ${\cal R}$.} Also listed
is the maximum expected flare class.}\label{tab:likelihood}
\begin{tabular}{lcrrrrr}
\hline 
$\log{(\cal R)}=$ & $<$3.0 & 3.0 & 3.5 & 4.0 & 4.5 & 5.0\\
\hline
$>$M1    &0{\%}&  2{\%}& 5{\%}& 12{\%}& 50{\%}& $\sim$80{\%}\\
$>$M3    &0{\%}&  $\sim$0{\%}& $<$1{\%}& 3{\%}& 20{\%}& 35{\%}\\
$>$X1    &0{\%}&  0{\%}& $\sim 0${\%}& $\sim$1{\%}& 10{\%}& 20{\%}\\
$>$X3    &0{\%}&  $0${\%}& $0${\%}& $\sim 0${\%}& 1{\%}& $1-2${\%}\\
\hline
Maximum: & $<$C9  &$<$M1 & $<$M4& $<$X1& $<$X4& $<$X10\\
\hline
\end{tabular}
\end{center}
\end{table}

Another example of a statistical study is that by
\refcite{schrijver2007}, who looks at the magnetic properties of
regions associated with almost 300 M- and X-class flares, and compares
that with the properties of 2,500 randomly selected active regions,
all within 45$^\circ$ of disk center. He notes that the regions with
large flares all have a pronounced
polarity-inversion line with a strong gradient across it. He
subsequently measures the unsigned flux ${\cal R}$ within about 15\,Mm
of such strong PILs (SPILs), and this is a good indicator for flares
that allows prediction of the largest flare to be expected from the
region (see Fig.~3 and 
Table~2; I return to the issue of predictability and the skill of
various metrics in Sect.~\ref{sec:instability}).

Based on the inspection of many full-disk SOHO-MDI magnetogram
sequences for regions with large flares, \refcite{schrijver2007}
argues that the SPILs appear to be associated with emerging flux, in
particular with emerging flux ropes (also supported by the detailed
field modeling by \barecite{schrijver+etal2007}; cf.\
Fig.~1). \refcite{welsch+li2008} show that there is a pronounced
tendency of increasing ${\cal R}$ to be correlated with increasing
total unsigned flux, i.e., with signs of flux emergence. Given the
resolution of full-disk MDI magnetograms and the likelihood that
emerging flux immediately cancels partly against pre-existing flux,
this is strong evidence for flux emergence as driver for the SPIL flux
metric ${\cal R}$. More evidence for the association with flux ropes
in particular is found when looking at shear at such SPILs.

%\section{Shear at the polarity-inversion line}
\label{sec:shear}
Shear at the polarity-inversion line is often associated with flaring activity
(examples are a statistical study by, e.g., \barecite{cui+eetal2007}, and
individual event studies by \barecite{kurokawa+etal2002}, 
\barecite{{brooks+etal2003}}, \barecite{{dun+etal2007}}, 
\barecite{schrijver+etal2007}, \barecite{okamoto+etal2008},
and \barecite{lites2008}; also 
coronal field studies described in Sect.~\ref{sec:overlying}).  The
same holds for eruptions:
\refcite{vrsnak+etal1991} report that the pitch angle of H$\alpha$
structures outlining filaments is telling regarding the eruption potential:
prominences with twist angles exceeding 50$^\circ$ and heights
exceeding 80\%\ of the footpoint separation erupted, while
those with angles below 35$^\circ$ and heights below 60\%\ of the
footpoint separation were stable. 

\refcite{manchester2007}, based on simulations of an emerging flux rope,
argues that shear motion at the polarity-inversion line in emerging flux is
a natural consequence of the Lorentz forces that develop as a rope
enters the corona. This may explain the general correspondence of field shear
and shear flows.

\section{Flare-related changes in the magnetic field}
\subsection{Before and after in the photosphere}\label{sec:beforeafter}
The conversion of large-scale electromagnetic energy in the process
of a flare or filament eruption changes the field configuration over
active regions, and this should in turn have consequences for 
the observable vector-magnetic field in the photosphere. Until recently,
these changes proved elusive (other than temporary changes
in line profiles during the flares, which are mostly if not
entirely a consequence of the
energy deposition in the near-atmospheric layers by populations 
of energetic particles generated early in the flare, see, e.g.,
\barecite{schrijver+etal2005c}). 

Part of the problem in finding flare-related field 
changes is that it cannot
really be done with vector-magnetic measurements based on
spectrograph scans, simply because these scans take too long to 
complete relative to the evolution of the photospheric field between
two successive magnetograms with spacings that are much longer than the
impulsive phase of the flare or eruption. For example, 
\refcite{dun+etal2007}, using a series of 12
vector-magnetograms in the course of five days of AR\,10486 (case
22 in Table~\ref{tab:regions}), find that magnetic shear angles, electric
current density, and current helicity increase at the sites of
impulsive flares at least a day prior to the main (X17 and X10)
flares. But they attribute these changes to the emergence of twisted
flux ropes rather than to changes associated with the flares per se.
Although they say their metrics decrease after the main flares, the
observational cadence is too low to unambiguously differentiate
flare-related relaxation of the field or the passage of flux ropes
through the photosphere prior to the post-flare magnetograms.

The analysis of line-of-sight magnetograms taken at much higher frequency 
has been successful in identifying differences in the magnetic
field just before and after major flares. 
\refcite{sudol+harvey2005}, for example, 
find: a) that the time scale for changes in the l.o.s.\ field is less
than $\sim 10$\,min., some even less than one minutes, but in all
cases abrupt; b) in many cases these changes are close to the noise
but in some cases very significant; c) the changes persist for at least
the two hours past the peak flare time studied by them, and in one
case for at least 5 hours. \referee{They note
that most of these observed field changes occur in spot
penumbrae, and that for the three flares for which TRACE coverage
is available, these changes occur at sites of flare ribbons in the
(E)UV. The authors argue that these field changes are lasting, and
thus likely not the result of atmospheric changes related to the
particle impact or heat conduction leading to the flare-ribbon formation;
perhaps the energy deposition occurs (very nearly) along the current-carrying
field lines involved in a reconnection event. The fact that 
these field changes are often close to the detection threshold even
for major flares will make it difficult for field modelers to reliably
deduce changes in the field energy from before to after flares in general
--~see Sect.~5).}

\refcite{wang+etal1994a} show cases in which the shear angle across the
polarity-inversion line of active regions increases during, and
persists after a set of five X-class flares, with a time resolution
that ranges from about 4-10\,min.\ in the best cases to about
45\,min. They propose as one possible interpretation that new flux
emerging just before the flare could be the cause of this increase,
but comment that it is hard to understand how that could change the
shear over more than 10,000\,km in such a short time.  On the other
hand, \refcite{chen+etal1994} report on shear studies of 18 regions around
M-class flares; they find no detectable changes, suggesting that
perhaps only the largest flares are associated with observable
changes.

Even very large flares and eruptions are associated with relatively
weak or very localized changes in the surface magnetic field. This
is, of course, part of the problem of measuring significant changes in
the field energy, helicity, and topology around the time of such
field destabilizations: too often, the field changes that are \referee{reported
are mixed in with} intrinsic changes in the photospheric field caused
by flows (horizontal and, often, vertical associated with flux emergence)
\referee{owing to the substantial time intervals between successive magnetograms,
thus masking  the changes associated with the flare or
eruption only. The potential of, e.g., the 1-min. cadence magnetogram
sequences for SOHO/MDI has not yet been utilized. Moreover, 
we shall need sensitive (vector-) magnetographs,
preferably for the chromospheric field, with high cadences and sizable
fields of view in order to learn about vector field changes at the lower
boundary at times of flares and eruptions.}

\subsection{Before and after in the corona}\label{sec:beforeafterabove}
There do not seem to be many studies that compare the coronal
appearance before and well after a major flare (well observed in,
e.g., case 24 in Table~1).  There are some
studies, however, that look at loops early in a flare/eruption and in
the late stages of the post-flare arcade. Among these is
\refcite{asai+etal2003}, who find that during the course of the
initial stages of the flare, the conjugate footpoints at first
indicate strongly sheared field lines, and later reveal less sheared
coronal connections.
\refcite{su+etal2007} study TRACE observations of 50 X- and M-class
two-ribbon flares, with well-defined, largely parallel ribbons. They
find that 86\%\ of these flares show a general
decrease in the shear angle between the main polarity inversion
line and pairs of conjugate bright ribbon kernels. 
They interpret this as a relaxation of the field towards a more potential
state because of the eruption that carries helicity/current with it, 
but one can readily argue that a similar decrease in shear angle would
be seen if sequentially
higher, less-sheared post-flare loops lit up with time as the loops
cool after reconnection. These results
are consequently ambiguous: they may show a decrease in shear, or
they may reflect that flares generally do not release all available
energy and part of the flux-rope configuration remains 
(cf., Sect.~\ref{sec:homology}).

\refcite{su+etal2007a}, from a study of 
a sample of 31 two-ribbon flares with CMEs, find that the primary
parameters describing the peak flare brightness and CME terminal
velocity are the total flux in the region, and the change in
shear angle between the early and late phases of the flare, and -~to
a significantly lesser degree~- the initial shear angle. The
change in shear angle may reflect a real decrease in shear and 
associated energy, but -~in view of the preceding paragraph~- may
also be a measure of the residual shear after the eruption; hence,
it remains unclear what the measured  shear angle is telling us in this case.

Compounding the efforts to understand the field evolution are the
relative roles of twist and writhe, and their coupling. One study on this
topic by
\refcite{romano+etal2003} reports that the number
of turns around an erupting filament initially is about 5, decreasing
to 1 later into the eruption (entry 25 in
Table~\ref{tab:regions}). Here we may see an exchange of twist and
writhe.

\section{The overlying coronal field}\label{sec:overlying}
%\subsection{Models}
Numerical MHD models continue to shed light on the potential causes of
the instability that leads to flares and, in particular, filament
eruptions, \referee{as always with the clear caveat that such modeling is
an approximating abstraction of reality}.  Some of these models rely
on the expected general slowness of reconnection in a solar
atmospheric plasma, except for particular field configurations,
specifically null points. The so-called breakout model, for example,
``exploits a vulnerability of multi-polar configurations, which
consist of two or more distinct flux systems separated by null points
in the corona, to rearrangements of the magnetic field's
connectivity'' (\barecite{devore+antiochos2008}).  Specifically, rapid
reconnection is anticipated as flux is pushed into a null region.  One
recent example of such fields is discussed by
\refcite{devore+antiochos2008}. They simulate a multi-polar
configuration with a null in the corona over a model active region
subjected to rotational flows (somewhat like rotating sunspots). Their
simulation run extends to cover three homologous eruptions in their
breakout concept.  These, by the way, are confined, in the sense that
they erupt only from the active region, but not into interplanetary
space.

Another concept is the 'tether-cutting' model (originally described by
\barecite{moore+roumeliotis1992}), in which (near-)photospheric
reconnection associated with flux emergence 
removes the 'anchors' to flux rope that can subsequently
reconnect with the overlying field, or even erupt. This is discussed
further in Sect.~\ref{sec:reconnection}.

A different possibility to destabilize a field configuration is
related to the kink and torus instabilities. The kink instability
occurs when the helicity, current, or winding-number for the field
in a flux rope exceeds a critical value. For example, 
\refcite{fan+gibson2004} perform line-tied MHD simulations
of an emerging flux rope into a pre-existing configuration. For a case
in which the twisted field in the top part of the rising rope has the
same general direction as the overlying arcade, they find that the
configuration becomes unstable after the coronal field reaches 1.76
full turns about the rope's axis. The erupting rope exchanges twist
for writhe as it evolves as is often -~but certainly not always~- seen
in filament eruptions.  In a case in which the rope's top-most winding
relative to the overlying arcade is reversed, no flux rope builds up
in the corona as reconnection proceeds from the very first emergence,
and consequently no instability develops.

The torus instability occurs whenever the gradient in the surrounding field
is strong enough, so that the forces exerted on a flux rope subject
to a perturbation cannot be contained by the overlying magnetic field. 
\refcite{kliem+toeroek2006} (also, e.g., \barecite{fan+gibson2007}) 
discuss the torus instability, and
find a critical field gradient beyond which the configuration is
unstable. They argue that line tying of the field to the photospheric
boundary in this case helps to stabilize a configuration, arguing that
it may even stabilize an emerging flux rope entirely until it is
semi-circular and thereafter may help to destabilize it.

\refcite{torok+kliem2007} discuss a unifying approach  
to slow and fast CMEs in which the torus instability drives the CME,
and in which the gradient in the surrounding field determines the
acceleration profile and eventual velocity of the ejecta: fast CMEs
for rapid decrease (as typical of ARs) and slow for gradual decrease
(as often over quiet Sun). \refcite{isenberg+forbes2007} start from a
similar field configuration as developed by
\refcite{titov+demoulin1999}, but  perform an analytical analysis
for eruptions without twist (i.e., without the subsurface current
introduced by
\barecite{titov+demoulin1999}).  They confirm the instability of the
configuration when the flux rope length exceeds a few times the
diameter of the flux rope.

\referee{Following} the proposition of the torus instability,
\refcite{wang+zhang2007}, based on potential-field models, find that
'the ratio of the magnetic flux in the low corona to that in the high
corona is systematically larger for the eruptive events than for the
confined ones'. 
\refcite{liu2008}, for a sample of ten flares, find 
a low field gradient for failed eruptions or kinking, and
a steeper gradient for successful eruptions. 
The power-law indices for the field decrease with
height for the failed eruptions and for the
kink and torus eruptions combined are $-1.62 \pm 0.05$ and $-1.91 \pm
0.15$, respectively, for height ranges from $\log(h[{\rm
km}])=3.7-4.7$. The use of potential-field models here is an
oversimplification and the findings may say more about the
distribution of the surrounding field on the photosphere than about
the actual field gradient in the low corona.

The above-mentioned studies deal with CMEs in which flux ropes that
may contain filaments are part of the story. It remains to be seen 
how this can apply to flares, or to eruptions with failed ejections,
but there is no obvious reason why the above arguments would not apply
to compact flares as much as they do to more extended flux-rope cum
filament systems. 

%\subsection{Observations}
In view of the theoretical and numerical considerations described in
the preceding subsection, it is no surprise that multiple
observational studies look at the coronal field configuration. Of
course, this brings us back to the problem that the coronal magnetic
field configuration cannot be known really well with present tools
(Sect.~\ref{sec:topology}). This is particularly bothersome because it
is clear that the coronal fields in flare-prone regions differ
markedly from the potential field based on the same vertical (or line
of sight) photospheric field; see, for example,
\refcite{schrijver+etal2005a}. This does not mean that currents
pervade much of the active-region corona, of course, but at least
points out that somewhere there must be currents strong enough to warp
the field relative to a pure potential field everywhere from the core
field to the peripheral field.  It appears that such currents survive
for typically only a day, while powering one or more flares during
their decay (e.g.,
\barecite{pevtsov+etal1994}; \barecite{burnette+etal2004}; 
\barecite{schrijver+etal2005a}).

How common is a strongly non-potential field?
\refcite{burnette+etal2004} compare
photospheric vector-magnetograms and coronal configurations as seen
by YOHKOH's SXT. They compute LFF models that best
match the coronal geometry, and find that over 90\%\ of the regions
studied are fitted well by such a model. 
\refcite{nindos+andrews2004} find that this is true for
only 60\%\ of their sample. Furthermore, \refcite{burnette+etal2004}
find that the $\alpha$ parameter derived from the vector-magnetogram
correlates well with that estimated for the coronal configuration.
Apparently, the currents that are responsible for the large-scale
non-potentiality of the coronal field close through the photosphere,
while small-scale structure of the $\alpha$ parameter in the
photosphere often does not contribute significantly to the large-scale
field configuration.

The electrical currents involved in non-potential regions
appear to emerge with the flux, according to several studies.
\refcite{pevtsov+etal2003a}, for example, use line-of-sight MDI magnetograms
and LFF field fits to EIT EUV images to estimate the helicity evolution in
emerging active regions. They find that the helicity parameter
$\alpha$ starts essentially at zero on first emergence for all six
regions that were well observed, and subsequently converges to a
maximum plateau value over the following 1.5\,d. They argue that this
means that the twist is in fact injected into the corona by 'the
spinning of the active region polarities driven by magnetic torque
from below'. Note that this means on first emergence, the coronal field
is apparently nearly potential, with $\alpha \sim 0$.
Work by \refcite{longcope+welsch2000A} supports this finding.

%Detailed comparisons of observations and models show that some filament 
%configurations are very nearly internally unstable (\barecite{bobra+etal2007}).

The studies in the literature, both observational and theoretical,
suggest that whether a flux rope erupts not at all, evolves into a
``failed eruption'', or makes it into the heliosphere as a CME depends
critically on the structure of the overlying field, in particular on
the gradient of the field with height, but likely also the direction
of the overlying field relative to the twist of the emerging flux
rope.

\begin{figure}
\begin{center}
\includegraphics*[width=0.3\textwidth]{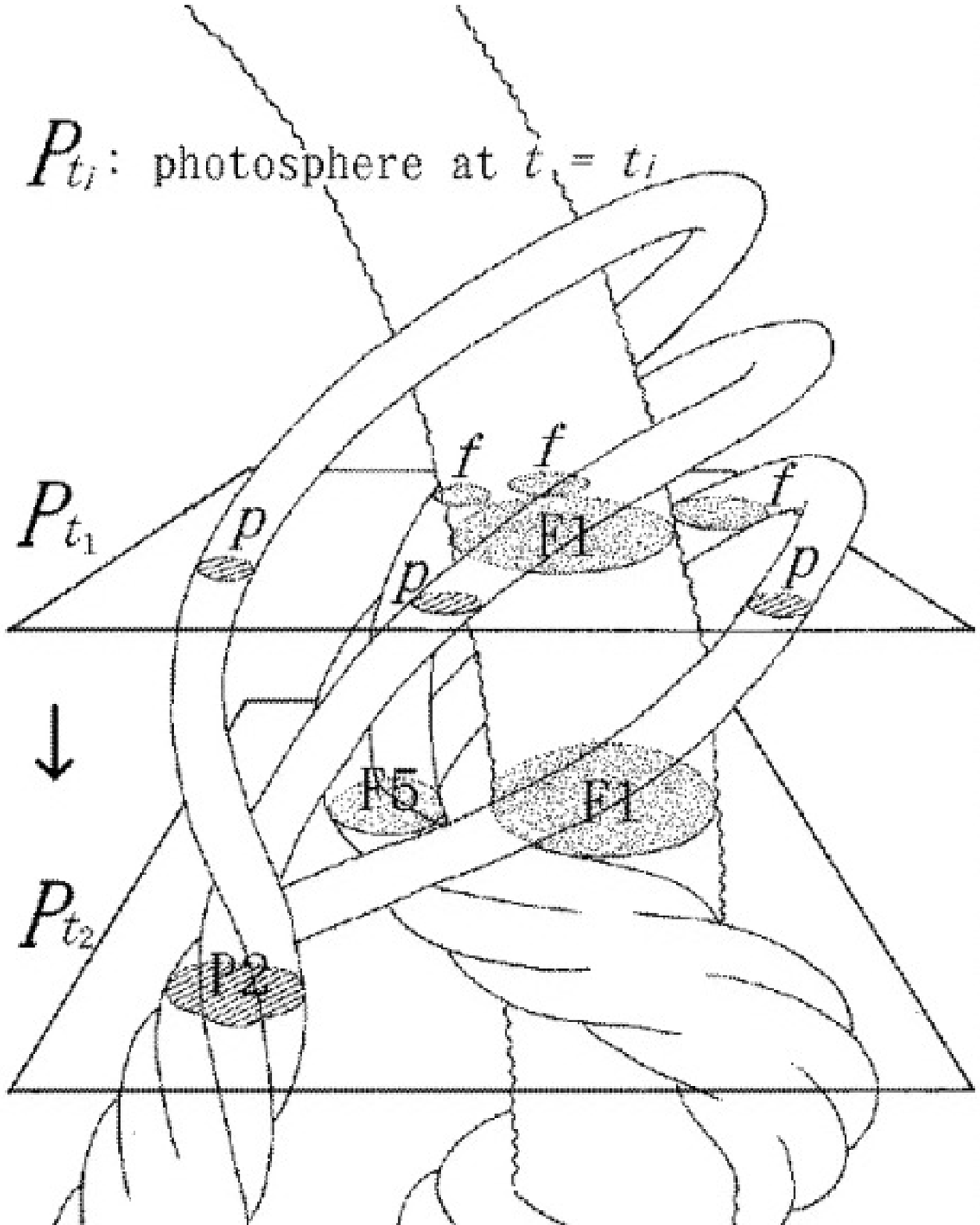}\includegraphics*[width=0.6\textwidth]{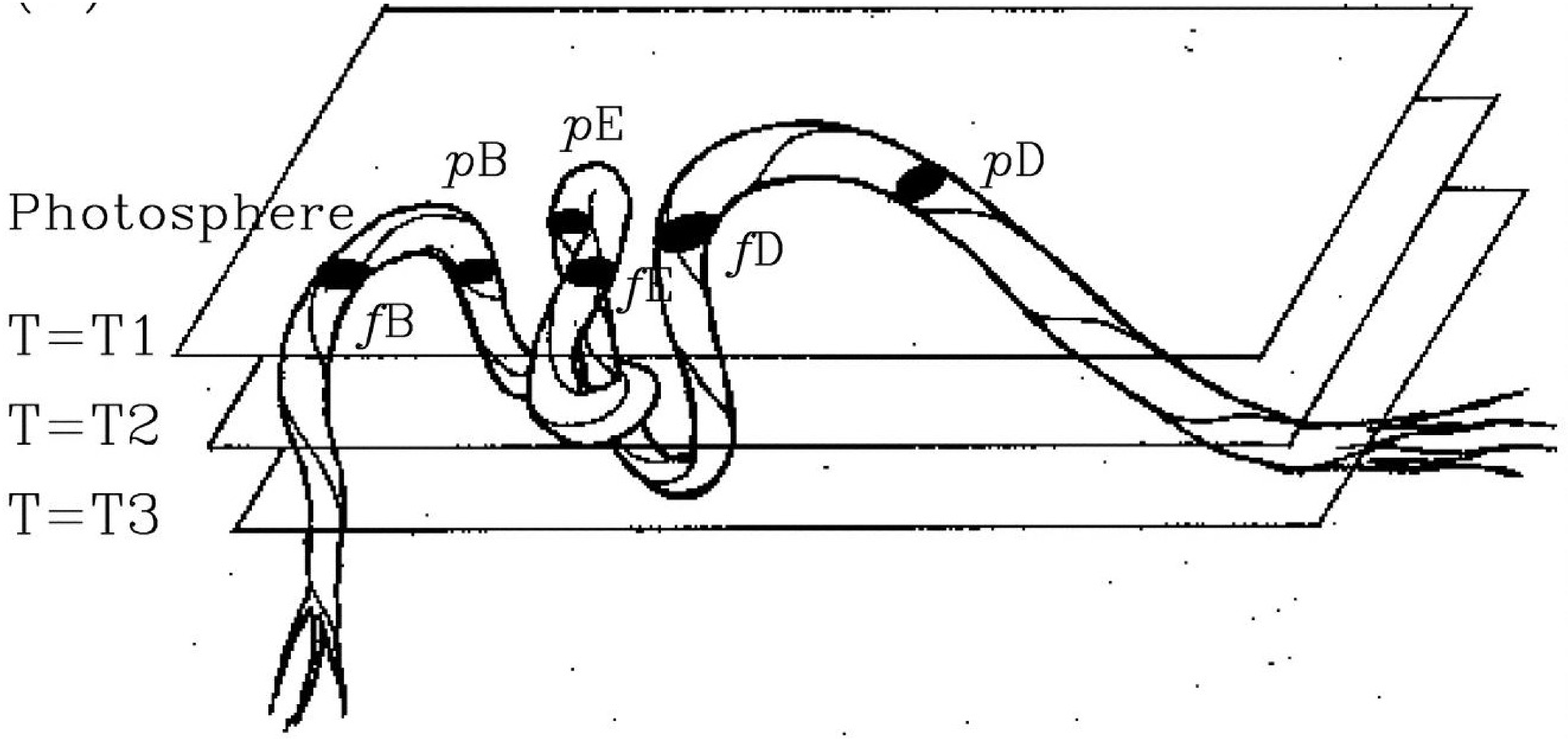}

\includegraphics*[width=0.4\textwidth]{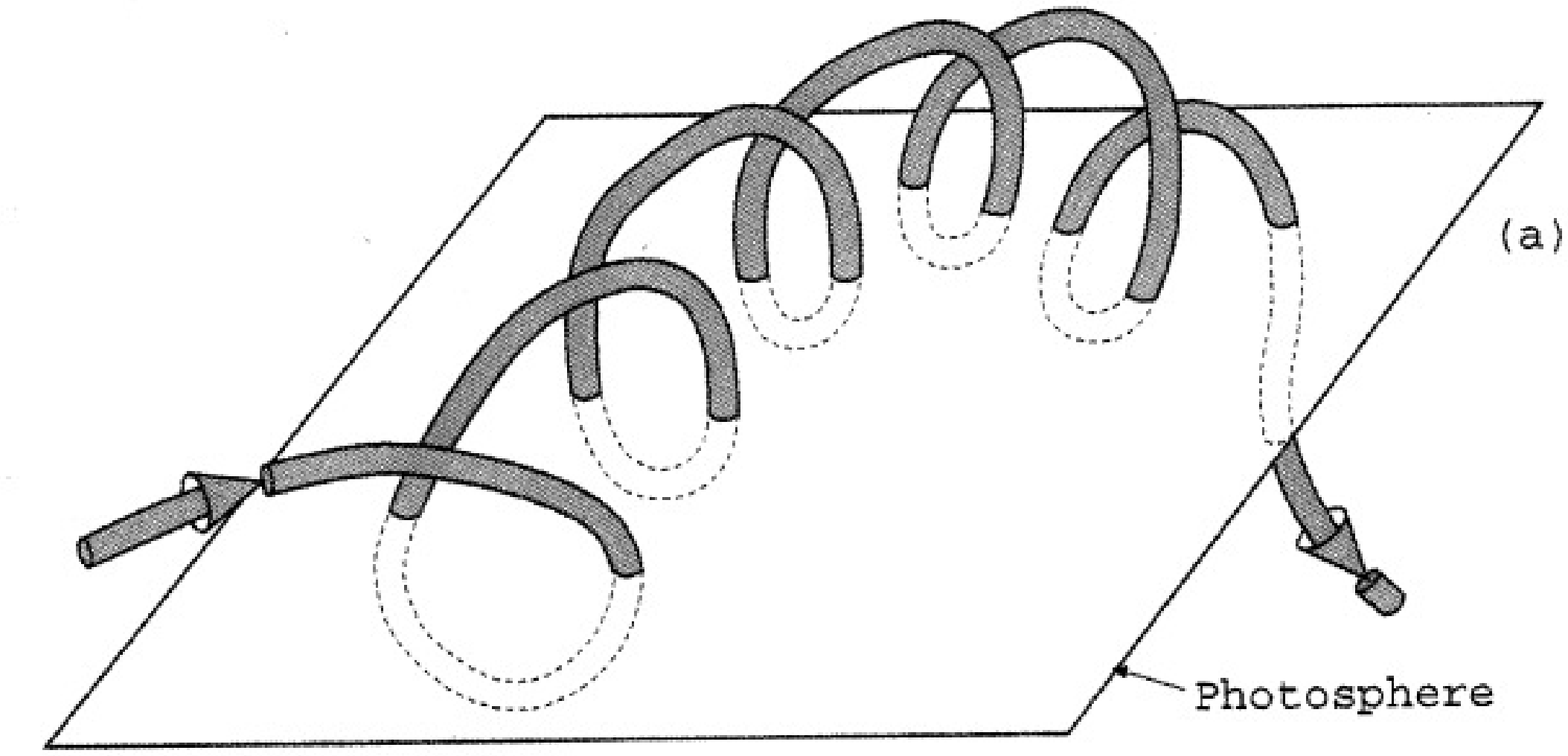}\includegraphics*[width=0.4\textwidth]{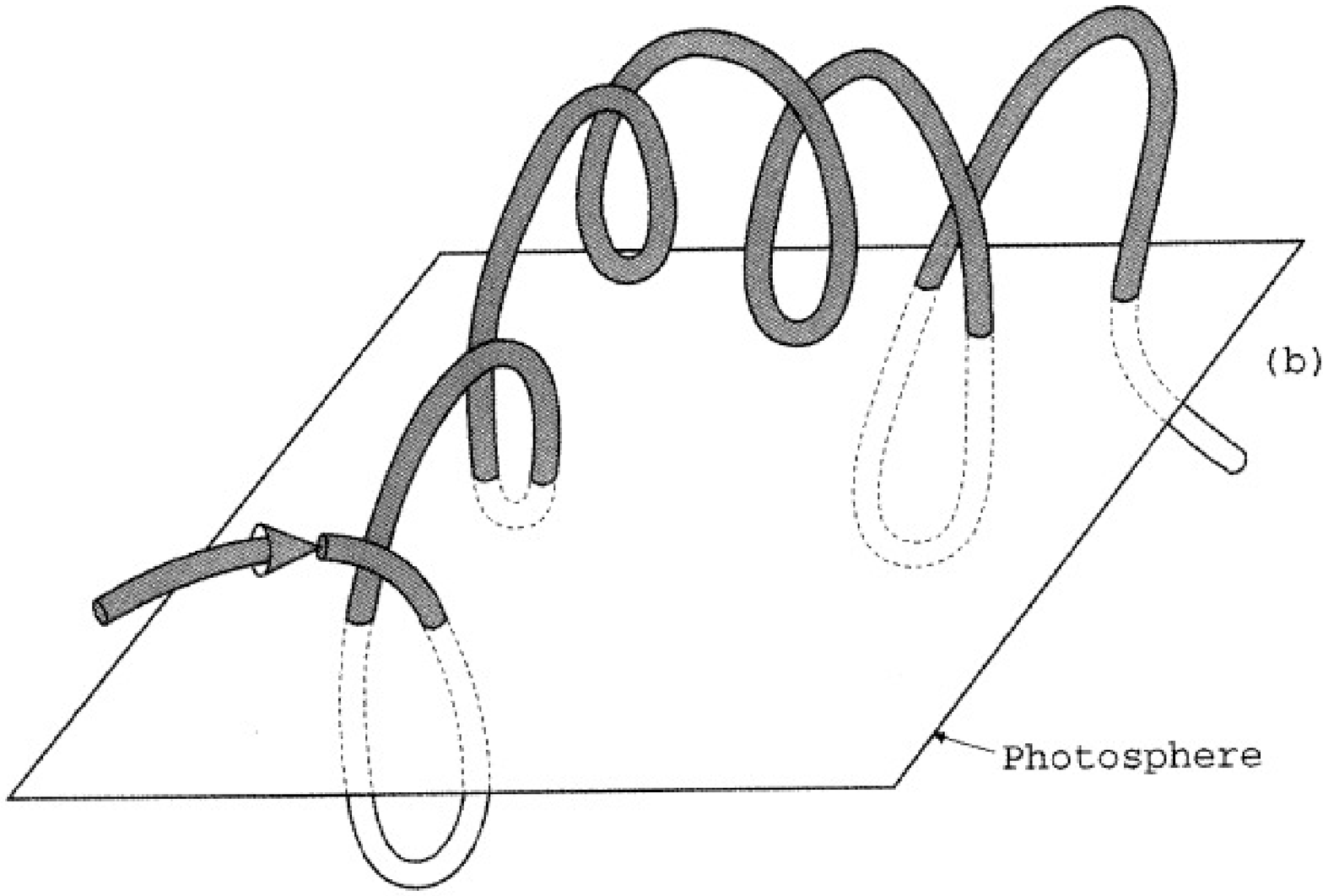}
\end{center} \caption{\label{fig:ishiikurokawa} Select examples of
proposed 3D configurations to explain the motions of pores around a
spot (top left), of the complex motions of multiple concentrations of
flux in an emerging region (top right), and the photospheric crossing
of an emerging flux rope (bottom two panels). In these sketches the
possibility of efficient supra-photospheric reconnection (as in the
Ellerman bombs of Fig.\ 5) has been ignored.  (Top left: from Ishii et
al., 1998; top right: Kurokawa et al., 2002; bottom: from Low, 2001)}
\end{figure}
\nocite{ishii+etal1998}\nocite{kurokawa+etal2002}\nocite{low2001} 

\section{Subsurface processes, emerging flux, and rotating spots}
\label{sec:subsurface} 
%\subsection{Observations}
Currents through the photosphere are often not neutralized when
integrating over either of the two polarities. \referee{It has been
argued (e.g., \barecite{melrose1991}, \barecite{melrose1995},
\barecite{wheatland2000}) that consequently these currents cannot have
been induced by non-current carrying flux emergence, because in that
case the currents should balance within each polarity.  Hence,
instead, at least some of these currents must have emerged from below
the photosphere.} This notion is compatible with the observational
evidence that the Sun's magnetic field is often twisted on emergence
(e.g., \barecite{kurokawa1987}, \barecite{tanaka1991} [see, e.g., his
Fig.\ 5], \barecite{leka+etal1996}, \barecite{vandriel+etal1997},
\barecite{ishii+etal1998}, \barecite{lopez+etal2000},
\barecite{okamoto+etal2008}).  \refcite{brooks+etal2003} study in
detail, with high spatio-temporal resolution, the conditions of two
emerging flux regions, one with a small flare, and one without, in the
field of view of the SVST. They find that the flaring region is
associated with the emergence of what looks like a twisted, i.e.,
current-carrying flux rope (compare Fig.~1), while the other has no
such appearance.

Not only emerging currents contribute to a region's free energy,
however.  Also systematic flow patterns at, or just below, the surface
can do so.  For example, \refcite{nightingale+etal2007},
report that rotating sunspots are associated with almost
all of the X-flares observed by TRACE since April 1998, and many of
the M-flares. 

It is difficult to differentiate twist in the emerging
flux from twisting flows. 
Support that these twists are associated with flows, at least below
the surface, is found by \refcite{komm+etal2005}, who estimate the
kinetic helicity density (${\bf v} \cdot {\bf \nabla} \times {\bf v}$)
below active regions using helioseismic measurements. They find that
the maximum helicity density in the outer 2\%\ of the Sun correlates
'remarkably well' with the X-ray flaring activity, perhaps -~as they
point out~- because of the link between helicity and twist in the
subsurface magnetic field.

Whereas sunspot rotation has been argued to add energy to the coronal
field, either because of the work done by the (sub-)photospheric flows
or because it reflects the emergence of a twisted, current-carrying
flux bundle, the study by \refcite{schrijver2007} shows that all X-M
flaring regions contain SPILs. As all X-flaring regions appear to
contain rotating sunspots, it may well be that these are related
phenomena, perhaps because many, if not all, flux bundles in a
flare-prone region have significant electrical currents associated
with them induced by preferential flow systems.

There appears to be a net relative helicity in each hemisphere, negative in the
north, positive in the south, corresponding to left-handed screws in
the north and right-handed ones in the south: the statistical study by
\refcite{pevtsov+etal1995} shows a 2:1 preferences of the sign of
helicity for the hemispheres, averaged over all latitudes and for two
cycles (see \barecite{longcope+pevtsov2003}). This twist, if existent
already below the surface, is likely to be affected by turbulent
convective buffeting of rising flux (see \barecite{lopez+etal2000},
for a discussion of an example of this). As
\refcite{longcope+pevtsov2003} argue, this would trade twist with
writhe of the emerging flux tubes, without affecting relative
helicity. The writhe should affect the tilt angle of dipole axes on
emergence, if not cause kinking of the subsurface flux arches into
what may be seen as complex, or even $\delta$-class regions. This
suggestion has been made, but I have not found this empirically
investigated.

%\subsection{Models}
On the theoretical side, a persistent problem is that even 
state-of-the-art numerical experiments still have difficulty in
successfully bringing flux bundles from some depth up into the
photosphere without them exploding in the strongly stratified
uppermost layers of the convective envelope (e.g.,
\barecite{fan2004}). Note that among the stabilizing forces in this
process we find increasing curvature, the Coriolis forces,
and Reynolds number
(e.g., \barecite{abbett+etal2000}, \barecite{abbett+etal2001},
\barecite{magara2006}, \barecite{cheung2006}). Models suggest that twist is
needed for the flux bundles -~thus turning them into flux ropes~- to
survive the rise. On the other hand, \referee{it appears that modelers
need to impose more twist than warranted by observations, as 
most active regions emerge with archfilament systems mostly aligned
with the eventual dipole axis of the region rather than nearly 
perpendicular to it as in  many simulations
(e.g., \barecite{archontis+etal2005}; \barecite{magara2006};
\barecite{manchester2007}; \barecite{fan2008}).
}

Starting
with too much twist subjects the emerging flux ropes 
to kink instabilities.
\refcite{fan2008} discusses models of rising 
flux tubes through a rotating medium stratified like a convective
envelope. She shows that if these non-axisymmetric flux tubes rise
with a twist of the observed hemispheric preference that is strong
enough to prevent substantial shredding during the rise, the twist
converts into a writhe that has the opposite sign than the observed
preferred tilt angle. If the twist is weakened by only a factor of
two this effect becomes weaker than that of the Coriolis force, and
the net result is a tilt that is comparable to the observed one. But
the consequence of this is that the rising 3D tube loses a lot of its
flux by shredding as it moves buoyantly through the surrounding medium.
In view of this, one sees this scenario develop: 
weakly twisted tubes explode and cause network field;
moderately twisted regions have a trail of shredded field behind them,
perhaps causing active-region nesting; and 
strongly twisted field should come
up coherently, perhaps subjected to a
kinking-instability leading to $\delta$-spots.

\section{The role of near-photospheric reconnection}\label{sec:reconnection}
With mounting evidence for the emergence of flux ropes as a
characteristic ingredient in major solar flaring, it is instructive to
look at some of the highest-resolution studies made of the process of
flux emergence.  These observations suggest that emerging magnetic
flux often does not emerge as a simple arch, as traditionally studied,
but rather as an undulating field that transits the photospheric layer
multiple times in sea-serpent fashion.  In such a geometry, the
magnetic field reaching into the chromosphere might drain its plasma
load into sub-photospheric dips in the field, with trapped material
that would not allow those field segments to rise. But perhaps
near-surface reconnection allows the heavily mass-loaded pockets of
magnetic field to be pinched off (thus creating a variant of the
tether-cutting concept in which the emerging and pre-existing flux ropes
are, in fact, one and the same).

\refcite{lites2008} discusses such 
observations of emerging flux made with Hinode's SP, NFI, and BFI. He
argues that the observations of opposite-polarity patches moving
towards each other, with downflows seen in the spectral data, 
present an opportunity for
plasma to drain back into the photosphere, and then be pinched off
from the supra-photospheric configuration, which can then rise into
the corona. Recent numerical simulations of emerging flux (e.g.,
\barecite{cheung+etal2008}) support such an interpretation.

\begin{figure}
\begin{center}
\includegraphics*[width=0.7\textwidth]{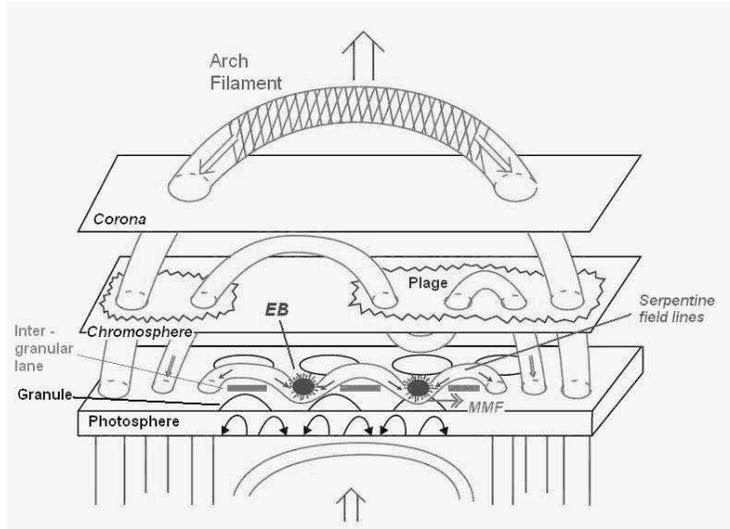}
\end{center}
\caption{\label{fig:schmieder}
Conceptual illustration (by Schmieder and Pariat, 2007) of
emerging flux, with undulating strands of flux with multiple
photospheric crossings. In this concept, the approach of pairs of
corresponding opposite-polarity patches on either side of a dipped
field line could lead to reconnection (in what might appear as
an Ellerman bomb, or EB) that pinches off the heavy sub-photospheric
segment, thus allowing the flux to rise. For field configurations
that can destabilize, twist and or kinking appears needed, as
illustrated in Figs.~\ref{fig:flarefluxmodel} and~\ref{fig:ishiikurokawa}.}
\end{figure}
The concept of pinching off the heavy, sub-photospheric dips in 
emerging, twisted magnetic field (such as sketched in the lower 
panels of Fig.~\ref{fig:ishiikurokawa}) has been suggested before
(also, e.g., \barecite{pariat+etal2004}). One example of
this is the interpretation of Ellerman bombs
as shown in Fig.~\ref{fig:schmieder}, taken 
from \refcite{schmieder+etal2007-pedia}.

\section{Instability, catastrophe, and predictability}\label{sec:instability}
Let me, finally, address two issues before ending this review. First a comment on the
nature of the instability. \referee{We have to ask
whether this instability can be the result of 
a bifurcation that occurs in certain conditions under which there
is more than one 
state for the field to be in given the observed photospheric
vector field. The existence of more than one field configuration for
the same lower boundary could be the explanation
for why lasting photospheric changes from pre- to
post-flare states are hard to find (although they have been reported on, see
Sect.~8.1).} Not much seems to have been written on this topic,
but interestingly, \refcite{isenberg+forbes2007} point out that they
'have found evidence that each T{\&}D [\barecite{titov+demoulin1999}]
equilibrium has a corresponding toroidal equilibrium with exactly the
same boundary conditions at the solar surface, which is unstable when
the T{\&}D equilibrium is stable and vice versa,' thus pointing to
a catastrophe scenario. Among the other
studies that address the existence of multiple solutions (some stable,
some unstable) I mention the catastrophe models by
\refcite{vanballegooijen+martens1989} 
and \refcite{forbes+isenberg1991}. Whether the possibility that
the nearly force-free state of the coronal field allows multiple
solutions to (very nearly) the same boundary condition will need
additional work in order to be able to assess these effects in the context
of field instabilities.

The second point relates to the predictability based on proposed
metrics for flare likelihood. A landmark study in this is the
work by
\refcite{barnes+leka2008} who evaluate the skill scores of 
forecasts based on four metrics: total flux, the ``total excess
energy'' (\barecite{leka+barnes2003a}), the SPIL flux metric ${\cal
R}$ (\barecite{schrijver2007}), and the ``effective connected magnetic
field'' (\barecite{georgoulis+rust2007}). They limit their sample
to regions within 30$^\circ$ from disk center to create a relatively
uniform set of observing conditions. 
They show that the highest
success rate for any of these single parameter, ${\cal R}$, is 92.2\%,
whereas a uniform forecast that nothing will every produce an M-class
flare or larger is 90.8\%. They conclude that none of these parameters
are robust as forecasting tools.

\section{In conclusion -- looking forward}
This, then, brings us back to the issue of ``understanding.'' If we
could deterministically forecast flares and eruptions, then we
certainly could claim success. But we might also have reached a deep
understanding if this were not the case: if flux-rope emergence is the
driver, or if self-organization occurs within the corona, then no such
thing as a ``deterministic forecast'' based on present-day observables
is feasible in principle.
How do we find out if either or both of these possibilities apply? It seems
that statistical studies for observations from interior to high corona
are essential. The relatively slow process of emergence and the
rapidity of flare onset also point to multi-day, 
high-cadence coverage for such coordinated observations.

These observations need to be complemented by numerical experiments of
flux emergence with a realistic treatment of the interior, of the
near-surface layers, and of the corona. Moverover, we shall need to
learn how to model the atmospheric field in order to
measure properties such as energy, helicity, and topology.

Coordinated multi-instrument observations of the Sun in the era of the
Solar Dynamics Observatory, supported by extensive spectroscopy and
numerical experiments, hold the promise of a breakthrough for our
understanding of magnetic instabilities involved in flares and CMEs.

\paragraph*{Acknowledgments.} I thank Hugh Hudson, 
Nariaki Nitta, and Alan Title for very helpful comments on early
drafts of this manuscript, and the referees for very constructive
suggestions to improve the presentation and to some key references.

%\bibliographystyle{klunamed}
%\bibliography{/net/star/Users/schryver/book/references/ref_karel} 

\end{document}